\documentstyle[12pt,aaspp]{article}
\def\ie{i.e.\ }
\def\eg{e.g.,\ }
\def\etal{et~al.\ }
\def\ltsima{$\; \buildrel < \over \sim \;$}
\def\simlt{\lower.5ex\hbox{\ltsima}}
\def\gtsima{$\; \buildrel > \over \sim \;$}
\def\simgt{\lower.5ex\hbox{\gtsima}}
\def\rquart{$r^{1\over 4}$}

\journalid{}{}
\articleid{}{}
\lefthead{Mihos \& Hernquist}
\righthead{Gasdynamics and Starbursts in Major Mergers}
\slugcomment{To appear in the {\it Astrophysical Journal}}

\begin{document}

\title{Gasdynamics and Starbursts in Major Mergers}

\author{J. Christopher Mihos\altaffilmark{1,2} and Lars
Hernquist\altaffilmark{3}}
\affil{Board of Studies in Astronomy and Astrophysics,\break
	 University of California, Santa Cruz, CA 95064\break
	 hos@pha.jhu.edu, lars@lick.ucsc.edu}

\altaffiltext{1}{Hubble Fellow}
\altaffiltext{2}{Current Address: Department of Physics \& Astronomy, Johns
Hopkins University, Baltimore, MD 21218}
\altaffiltext{3}{Alfred P. Sloan Foundation Fellow, Presidential Faculty
Fellow}

\begin{abstract}

Using numerical simulation, we study the development of gaseous inflows
and triggering of starburst activity in mergers of comparable-mass disk
galaxies. Our models cover a range of orbits and internal structures for
the merging galaxies. In all encounters studied, the galaxies experience
strong gaseous inflows and, using a density-dependent Schmidt law to
model star formation, moderate to intense starburst activity. We
find that galaxy structure plays a dominant role in regulating activity.
The gaseous inflows are strongest when galaxies with dense central bulges
are in the final stages of merging, while inflows in bulgeless galaxies are
weaker and occur earlier in the interaction. Orbital geometry plays only
a relatively modest role in the onset of collisionally-induced activity.
Through an analysis of the torques acting on the gas, we show that these
inflows are generally driven by gravitational torques from the host galaxy
(rather than the companion), and that dense bulges act to stabilize galaxies
against bar modes and inflow until the galaxies merge, at which point rapidly
varying gravitational torques drive strong dissipation and inflow of gas
in the merging pair. The strongest inflows (and associated starburst
activity) develop in co-planar encounters, while the activity in inclined
mergers is somewhat less intense and occurs slightly later during the
merger. To the extent that a Schmidt law is a reasonable description of
star formation in these systems, the starbursts which develop in mergers
of galaxies with central bulges represent an increase in the star formation
rate of two orders of magnitude over that in isolated galaxies. We find that
the gaseous and stellar morphology and star-forming properties of these
systems provide a good match to those of observed ultraluminous infrared
galaxies. Our results imply that the internal structure of the merging
galaxies, rather than orbital geometry, may be the key factor in producing
ultraluminous infrared galaxies.

\end{abstract}

\keywords{galaxies:interactions, galaxies:starburst, galaxies:active,
galaxies:evolution, galaxies:structure}

\vfil\eject

\section{Introduction}

Collisions and mergers of galaxies are thought to play a crucial role
in the onset of certain unusual phenomena in galaxies; specifically
starbursts and the occurrence of nuclear activity. The optical and
infrared colors of many peculiar galaxies can be understood in terms of bursts
of star formation triggered by galaxy encounters (Larson \& Tinsley 1978;
Joseph \etal 1984).  Colliding galaxies often display higher levels of
H$\alpha$ emission (\eg Kennicutt \etal 1987; Bushouse 1987), radio continuum
emission (Hummel 1981; Condon \etal 1982) and infrared emission (Lonsdale,
Persson, \& Matthews 1984; Solomon \& Sage 1988) than isolated galaxies,
indicative of strong star formation activity (\eg Joseph \& Wright
1985; Joseph 1990). Many of the far infrared luminous galaxies
identified by IRAS are found to be undergoing collisions (\eg Soifer \etal
1984ab; Sanders \etal 1986, 1988ab; Armus \etal 1987; Kleinmann \etal
1988). Furthermore, the fraction of interacting systems found in IRAS
selected samples increases with luminosity (Lawrence \etal 1989),
suggesting that collisions play a major role in triggering powerful
starbursts. In fact, the brightest infrared-luminous objects, with quasar-like
infrared luminosities of $L_{FIR} > 10^{12}$ L$_{\sun}$, all show
morphological peculiarities indicative of encounters, such as multiple
nuclei and tidal features (Sanders \etal 1988a, Melnick \& Mirabel 1990).
Millimeter CO observations of these galaxies have revealed massive pools of
molecular gas in their central regions (\eg Sanders \etal 1987; Sargent
\& Scoville 1991; Scoville \etal 1991), providing fuel for starbursts or
ultimately perhaps AGN activity. However, not all interacting systems show
elevated tracers of star formation (Keel \etal 1985; Bushouse 1986; Kennicutt
\etal 1987), raising questions about the detailed nature of the starburst
triggering mechanism.

Evidence also exists linking collisions and mergers to the onset of nuclear
activity in galaxies (see Stockton 1990; Barnes \& Hernquist 1992a,
Osterbrock 1993 for reviews). A variety of studies suggest that Seyfert
galaxies are preferentially found in interacting systems (\eg Dahari 1984,
Kennicutt \& Keel 1984; MacKenty 1989).  Investigations of the environments
of quasar host galaxies also reveal an excess of companion galaxies (McLeod
\& Rieke 1994; Hutchings, Crampton, \& Johnson 1995; Bahcall, Kirhakos,
\& Schneider 1995a), and many Seyferts, quasars, and radio galaxies exhibit
signs of encounters and mergers in the form of tidal debris and multiple
nuclei (van Albada \etal 1982; Smith \& Heckman 1989ab; MacKenty 1990;
Hutchings \& Neff 1992; Armus \etal 1994; Bahcall, Kirhakos, \& Schneider
1995b). In fact, the most luminous IR galaxies may actually harbor buried
AGNs (\eg Sanders \etal 1988a; Leech \etal 1989; Veilleux \etal 1995),
suggesting a possible evolutionary link between mergers, luminous IR galaxies,
and quasars (\eg Sanders \etal 1990).

It is clear from the observations that gasdynamics and star formation
in mergers must be responsible for much of the behavior associated with
active galaxies. In addition, these processes may also be quite
important for interpreting many aspects of ordinary galaxies. As
one example, the ``merger hypothesis" for the origin of early-type galaxies
argues that present-day ellipticals form primarily from mergers of disk
galaxies (Toomre 1977; Schweizer 1982, 1990). The tendency for mergers
to produce remnants possessing \rquart\ surface brightness profiles
has been well-demonstrated both observationally (\eg Wright \etal 1990;
Stanford \& Bushouse 1991) and theoretically (\eg Barnes 1988, 1992; Hernquist
1992, 1993a). Furthermore, many ellipticals exhibit low surface brightness
loops and shells (Malin \& Carter 1980; Schweizer \& Seitzer 1988) which
are thought to form naturally during disk galaxy mergers (Hernquist \&
Spergel 1992; Hibbard \& Mihos 1995). However, the merger hypothesis in
its simplest form suffers from a serious defect: the central phase space
density in stellar disk galaxies is much lower than the observed central
phase space density of ellipticals, so mergers of pure stellar disks
produce remnants less centrally-concentrated in phase space than real
ellipticals (Carlberg 1986; Gunn 1987; Hernquist, Spergel, \& Heyl 1993).
In principal, this objection can be overcome if gaseous dissipation and a
starburst during a merger acts to increase the central stellar density of
the remnant (Kormendy \& Sanders 1992).

The role played by merger-driven gasdynamics and starbursts in shaping normal
galaxies may be manifested in other ways as well. The kinematically distinct
cores observed in many elliptical galaxies (\eg Kormendy 1984; Franx \&
Illingworth 1988; Bender 1990ab; Forbes, Franx, \& Illingworth 1995) may
be formed through dissipation and subsequent central starbursts (Hernquist
\& Barnes 1991; Mihos \& Hernquist 1994c). Such processes may also give rise to
gradients in color profiles (Franx, Illingworth, \& Heckman 1989; Peletier
\etal 1990), metallicity profiles (Bender \& Surma 1992; Davies \etal 1993;
Mihos \& Hernquist 1994d; Surma \& Bender 1995), and the mean age of the
stellar populations in galaxies. Further from the galaxy center, induced
star formation may lead to the formation of young globular clusters (\eg
Ashman \& Zepf 1992; Whitmore \etal 1993) and perhaps even dwarf irregular
galaxies in tidal tails (Zwicky 1956; Schweizer 1978; Mirabel, Dottori, \&
Lutz 1992; Barnes \& Hernquist 1992b Elmegreen \etal 1993). Minor mergers
may also ``sweep" galactic disks clean of cold gas, transforming late-type
disk galaxies into earlier Sa or S0 galaxies (Hernquist 1989;
Mihos \& Hernquist 1994a). Therefore, any attempt to model the
relationship between galaxy collisions and violent activity or galaxy
evolution must ultimately account for the effects of dissipation and star
formation.

Theoretical work on these problems is relatively new. Early attempts to
include gasdynamics in mergers utilized a ``sticky particle" approach
for the evolution of the gas (\eg Negroponte \& White 1983;
Noguchi 1988). More sophisticated techniques employed
smoothed particle hydrodynamics to follow the evolution of the ISM
in merging galaxies (Barnes \& Hernquist 1991, 1995), giving
a more complete description of the physical conditions in the
gaseous components of the galaxies. These studies reinforce the notion
that galaxy mergers can drive significant inflows of gas under a wide
range of conditions. The first calculations to include star formation
and gas depletion in merging galaxies (\eg Noguchi \& Ishibashi
1986; Noguchi 1991; Olson \& Kwan 1990ab; Mihos \etal 1992, 1993; Mihos
1992) indicated that star formation rates can be raised by more than
an order of magnitude during a merger, and that these starbursts can
significantly deplete the galaxies of cold gas. However, these later
star-forming simulations all employed sticky particle hydrodynamics
and relatively crude galaxy models, and, moreover, suffered from
limited kinematic and spatial resolution due to the relatively small
number of particles used in the simulations (typically $N \sim 1-3\times
10^4$).

Here we expand on previous studies to assess more generally the nature
of merger-induced gas inflows and starbursts, and to make predictions which
may be tested observationally. We employ a more physical description of
the relevant physics and employ galaxy models which are more realistic than
those which have been used previously. In particular, we incorporate a
prescription for star formation into the hydrodynamical model for the ISM
which allows us to follow the full evolution of the merging galaxies,
including the production of a young starburst population. Furthermore, we
use fully self-consistent galaxy models, with structural properties chosen
to mimic the observed properties of nearby disk galaxies. These improvements
enable us to investigate the detailed dynamics responsible for driving
inflow and starburst activity during galaxy mergers.

In this paper we focus on ``major mergers" -- mergers involving
comparable mass galaxies -- in view of their likely relevance to
ultraluminous infrared galaxies and the formation of elliptical galaxies.
We survey mergers of pairs of disk galaxies having dark matter halos
and, optionally, dense compact bulges. In each case considered, we find
that the interaction and subsequent merger initiates a strong inflow of
gas into the nucleus of the remnant, triggering intense but short-lived
bursts of star formation. We isolate the physics responsible for these
inflows and demonstrate that their history depends sensitively on the
structure of the progenitor galaxies. We also investigate the effects of
orbital geometry by varying the inclination of the merging disks and find
that starbursts occur generically in encounters that lead to rapid merging.

We compare our results to those of previous theoretical work and
consider their relation to observations of starbursting and active
galaxies. In particular, we show that the emission from the
ultraluminous infrared galaxies can be plausibly explained by merger-induced
starbursts, without invoking a non-thermal energy source (\ie an AGN).
Although our models cannot follow the evolution of the nuclear gas on
parsec scales, if dissipation continues to act on the gas in those
regions, our results may also be applicable to the triggering and
fueling of active nuclei in galaxies. Finally, we examine how our results
depend on the assumptions and limitations inherent to our models, and
summarize questions for future interest.

\section{Numerical Technique}

Modeling the full evolution of galaxy mergers involves many distinct
physical processes, including gravitational dynamics, the hydrodynamics of
the ISM, and star formation. To follow these coupled effects, the simulations
described in this paper were performed using a hybrid N-body/hydrodynamics
code known as TREESPH (Hernquist \& Katz 1989), modified to include star
formation in the gas (Mihos \& Hernquist 1994e). We include here a brief
description of the numerical techniques; for a more thorough discussion, we
refer the reader to the above references.

\subsection{TREESPH}

To model composite systems of gas, stars, and dark matter, TREESPH uses a
hierarchical tree method (Barnes \& Hut 1986; Hernquist 1987, 1990b) to
calculate gravitational forces, and smoothed particle hydrodynamics (SPH;
Lucy 1977, Gingold \& Monaghan 1977) to follow the evolution of the gas.
Treecodes offer the best compromise for studies of merging galaxies, as
they place no constraints on the symmetry of the system or on spatial
resolution (other than that set by the gravitational smoothing length), while
offering a very efficient $N\log N$ scaling of computing time. Gravitational
forces are calculated using a tolerance parameter $\theta=0.7$ and including
terms up to quadrupole order in the multipole expansions. A cubic spline is
used to soften gravitational forces (Goodman \& Hernquist 1991), with
different softening lengths for different species of particles. In the models
described here, the softening lengths of the disk, bulge, and halo particles
are $\epsilon_d=0.08, \epsilon_b=0.06,$ and $\epsilon_h=0.37$, respectively.

SPH is a Lagrangian technique in which the gas is partitioned into fluid
elements represented by particles. These particles obey equations of motion
similar to those for the collisionless particles, but with additional terms
to describe pressure gradients, viscous forces, and radiative effects in
the gas. The SPH particles have individual timesteps chosen to satisfy the
Courant condition (Monaghan 1992) with Courant number $C=0.5$. A conventional
form of the artificial viscosity is used (\eg Monaghan \& Gingold 1983;
Monaghan \& Lattanzia 1985), with parameters $\alpha=0.5, \ \beta=1.0$.

To optimize the spatial dynamic range of the code, SPH particles have
their own smoothing lengths, such that a constant number ${\cal N}_s$
of neighbors are contained within two smoothing lengths. Estimates
of the hydrodynamic variables are symmetrized to preserve momentum
conservation, as described by Hernquist \& Katz (1989). The choice
of ${\cal N}_s$ represents a compromise between the goal of resolving
smaller scales (small ${\cal N}_s$) and improved accuracy in the
smoothed quantities (large ${\cal N}_s$). To reduce error propagation,
we have chosen to smooth the hydrodynamic properties over ${\cal N}_s=96$
neighbors. We note, however, that our results are in good agreement with
earlier calculations which employed ${\cal N}_s=30$ (\eg Barnes \& Hernquist
1991; Hernquist \& Barnes 1994), indicating that the results are rather
insensitive to the choice of ${\cal N}_s$.

We have opted to employ an isothermal equation of state for the gas with
a temperature $T_{gas} = 10^4$ K. Owing to limitations imposed by mass
resolution with a finite number of particles, we are unable to describe
a multiphase ISM (\eg McKee \& Ostriker 1977). As a result, earlier models
which explicitly followed the radiative heating and cooling of the gas
(\eg Hernquist 1989) employed a cutoff temperature in the cooling curve
at $T_c=10^4$ K. Because of the short cooling time of disk gas, fluctuations
in the gas temperature are quickly radiated away, so that most of the gas
resides near this cutoff temperature. As a result, simulations with an
isothermal equation of state differ little from those employing more
``realistic" ones (Barnes \& Hernquist 1995). Previous simulations have
also shown that the hydrodynamic evolution of the ISM in mergers between
disks and dwarf companions are quite robust against moderate changes in
the equation of state (Hernquist \& Mihos 1995). By choosing an isothermal
equation of state for these models, we expedite the calculation with little
cost in accuracy, and simplify the interpretation of the results. However,
the current models are insufficient to study fragmentation and star formation
in tidal tails, as our models neglect adiabatic cooling which is important
for rapidly expanding gas in the tidal debris.

The equations of motion are integrated using a time-centered leap frog
algorithm (\eg Press \etal 1986). For the collisionless particles, a
fixed time step of $\Delta t=0.16$ is used for the bulgeless galaxies and
$\Delta t=0.08$ for galaxies which possess bulges. Scaled to values
appropriate for the Milky Way (see \S2.3 below), these choices correspond
to $2\times 10^6$ years and $10^6$ years, respectively. For our fiducial
mergers, these timesteps are reduced by a factor of three after the dense
gas cores develop in the galaxies' centers, to improve energy conservation.
The SPH particles are allowed to have timesteps smaller than the collisionless
particles by powers of 2 in order to satisfy the Courant condition (Hernquist
\& Katz 1989).  Due to the use of an isothermal equation of state, and the
effects of star formation (see \S 2.2 below), energy is not strictly conserved
in these calculations. However, based on previous calculations (Barnes
\& Hernquist 1995; Hernquist \& Mihos 1995), integration errors are
responsible for drifts in the total energy of only $\sim$ 0.1\% for
the parameters employed.

\subsection{Star Formation Algorithms}

To model star formation, we use a variant of the Schmidt law (Schmidt 1959),
which relates the star formation rate in disk galaxies to the local gas
density by SFR (M$_{\sun}$ yr$^{-1}$ pc$^{-3}$) $\propto \rho_{gas}^n$,
where $n$ has been determined empirically to lie in the range $1\simlt
n\simlt2.5$ (see Berkhuijsen 1977; Kennicutt 1989 and references therein).
Retaining the Lagrangian nature of SPH, we parameterize the star formation
rate per unit mass in a gas particle according to $\dot M_{gas}/M_{gas} =
C\times\rho_{gas}^{1 \over 2}$. Averaged over volume, this prescription
yields an index $n=1.5$ in the classical Schmidt law (Mihos \& Hernquist
1994e). The constant of proportionality, $C$, is chosen such that an
isolated disk galaxy forms stars at a rate of $\sim$ 1 M$_{\sun}$ yr$^{-1}$,
yielding a depletion time for the disk of $\sim$ 5.5 Gyr, similar to values
inferred for present day disk galaxies (Kennicutt 1983).

We include the injection of energy into the ISM due to supernovae and stellar
winds from massive stars (``feedback") by imparting kinetic energy to the
surrounding ISM. At each timestep, the total amount of energy released via
massive star evolution $E_{SN}$ is calculated from the star formation rate
in each SPH particle, and a fraction $\varepsilon_{kin}$ of this energy
is used to provide a radial ``kick" to neighboring SPH particles. The
magnitude of this kick is given by $\Delta v_i = \sqrt{2 (w_i \varepsilon_{kin}
E_{SN}) / M_i}$, where $i$ refers the neighbor being perturbed and
$w_i$ is an energy weighting based on the smoothing kernel (see Mihos
\& Hernquist 1994e). In order to limit the amount of kinetic energy
injected into the ISM via star formation, we constrain $\varepsilon_{kin}$
so that the ISM in an isolated disk model maintains a constant vertical
scale height with time. Tests employing isolated disk models indicate that
$\varepsilon_{kin}=10^{-4}$ provides good results in terms of the scale
height of the disk gas and distribution of star formation (Mihos \& Hernquist
1994e).  Using $\varepsilon_{kin}=10^{-4}$ and $E_{SN}=10^{51}$ ergs,
$\Delta v_i$ is typically $\simlt$ 0.1 km s$^{-1}$ for nearby neighbors,
so that the {\it total} velocity perturbation in each star-forming particle
is typically $\sim$ 1 km s$^{-1}$. Concurrent with the energy release, star
formation injects metals into the ISM, allowing us to track metallicity
enrichment and to explore the development of gradients in the merger remnant.

To describe the effects of gas depletion and formation of a young stellar
population, we employ a technique involving ``hybrid" SPH/young star
particles (Mihos \& Hernquist 1994e) which are gradually converted from
gaseous to collisionless form. We characterize these particles by both a
total mass and a gas mass; the gravitational forces on and acceleration of
a hybrid particle are calculated using its total mass, while the gas mass
is used to calculate the hydrodynamic forces and properties of the gas.
Through star formation, the gas mass of a hybrid particle is reduced, while
its total mass remains fixed, thereby mimicking the effects of gas depletion.
When the gas mass fraction of a hybrid particle drops below 5\%, it is
converted to a collisionless particle, with its remaining gas mass smoothed
out among its nearest neighbors. These converted particles thus act as a
tracer of the young stellar population formed in a starburst event.

Our method of using hybrid particles to treat star formation and gas
depletion has a number of computational advantages. The total number
of particles remains fixed, eliminating the need to invest computational
resources in evolving large numbers ($N>10^7$) of young star particles.
Unlike methods which convert gas particles {\it in toto} (\ie Summers
1993), our algorithm places no constraint on the conversion mass involved
in a star formation event, allowing the models to describe the mass and
time scales necessary to examine the detailed star forming properties
of the system. The main disadvantage to our approach is that it assumes
the newly formed stars and the parent gas are dynamically coupled until
the gas is fully depleted. In reality, the two components will evolve
separately, as the gas can experience shocks and dissipation, while the
stars will evolve in a collisionless manner. This discrepancy will be
greatest if, \eg a shock passes through hybrid particles composed of
roughly equal amounts of gas and stars. However, this discrepancy will be
small if the hybrid particles are predominantly in one phase or the other,
as is the case in starburst events, where gas is converted to stars over
short timescales (\ie less than the dynamical timescale, $\sim$ 100 Myr).
We believe, therefore, that the mass-weighted error introduced by our
method will be relatively small for the simulations presented here.

\subsection{Galaxy Models}

Our galaxy models are constructed using the technique
described by Hernquist (1993b). The galaxies consist of a spherical ``dark''
halo, an exponential disk comprised of both stars and gas, and, optionally,
a compact central bulge. The halo is modeled as a truncated isothermal
sphere, whose density is given by
$$
     \rho _h (r) \, = \, {{M_h}\over{2\pi ^{3/2}}} {{\alpha}\over
      {r_c}} \, {{\exp (-r^2/{r_c^2})}\over{r^2+\gamma^2}} \, ,
$$
where $M_h$ is the halo mass, $r_c$ serves as a cut--off radius,
$\gamma$ is a ``core'' radius, and $\alpha$ is a normalization
constant defined by
$$
     \alpha \, = \, \left [1 \, - \, \sqrt{\pi} q \exp (q^2)
      \left ( 1 - {\rm erf} (q) \right ) \right ] ^{-1} \, ,
$$
where $q=\gamma / r_c$. The exponential stellar disks follow the
density profile
$$
     \rho _d (R,z) \, = \, {{M_d}\over{4\pi h^2 z_0}} \, \exp (-R/h)
      \, {\rm sech}^2 \left ( {{z\over{z_0}}} \right ) \, ,
$$
where $M_d$ is the disk mass, $h$ is a radial scale--length, and
$z_0$ is a vertical scale--thickness. The gas disks follow the same
radial profile as the stellar disks, but with a smaller vertical
scale--thickness, dependent on the gas temperature and galactic radius.
In some models we also include bulges, whose density
is given by an oblate generalization of the potential-density pair
of Hernquist (1990a) for spherical galaxies and bulges:
$$
     \rho _b (m) = {{M_b}\over{2\pi a c^2}} {1\over{m(1+m)^3}}
$$
where $M_b$ is the bulge mass, $a$ is a scale--length along the major
axis, $c$ is a scale--length along the minor axis, and
$$
     m^2 = {{x^2 +y^2}\over{a^2}} + {{z^2}\over{c^2}} \cdot
$$
Particles are distributed in space according to the density profiles
outlined above, with velocities initialized using moments of the Vlasov
equation and approximating the velocity distributions by Gaussians.

In the models described here, we use a system of units in which the
gravitational constant $G=1$, the disk mass $M_d=1$, and the radial
scale length of the disk $h=1$. Scaling these values to those typical
of the Milky Way (Bahcall \& Soneira 1980; Caldwell \& Ostriker 1981),
unit length is 3.5 kpc, unit mass is $5.6\times 10^{10}$ M$_{\sun}$,
unit velocity is 262 km/s, and unit time is $\sim$ 13 million years.

The galaxies are all characterized by halo and disk parameters
$M_h=5.8$, $\gamma=1.0$, $r_c=10.0$, and $z_0=0.2$. The stellar velocity
dispersion in the disk is such that the Toomre Q parameter is normalized
to 1.5 at the solar radius ($R_{\sun}=8.5/3.5$), and varies only weakly
with radius. In the models which include bulges, these bulges have mass
$M_b=1/3$ and scale--lengths $a=0.2$ and $c=0.1$. The numbers of collisionless
particles associated with each component are $N_h=32768$, $N_d=32768$, and
$N_b=8192$.  The disk gas, comprising 10\% of the total disk mass, is
represented by 16384 hybrid SPH particles with an isothermal gas temperature
of $T_{gas}=10^4$ K.

\section{Fiducial Models}

In \S3 we concentrate on the evolution and star-forming properties of
two fiducial galaxy mergers which differ only in the structure
of the progenitor galaxies. We investigate the effect of varying encounter
geometry in a wider, but less in-depth analysis of merger simulations in
\S4.

Our fiducial models involve two identical, equal-mass disk galaxies colliding
on a nearly parabolic orbit which leads to rapid merging over a few disk
rotation periods. The galaxies are initially placed at a separation
of $r_{init}=30$, and at first close passage the pericentric separation
(for the ideal Keplerian orbit) would be $r_{peri}=2.5.$ The geometry
of the encounter is described by the inclination of the disks relative to the
orbital plane ($i_1$ and $i_2$), and their argument of periapse ($\omega_1$
and $\omega_2$) (see, \eg Toomre \& Toomre 1972). Our fiducial mergers
involve one exactly prograde galaxy ($i_1=0\deg$, $\omega_1$ undefined) and
one highly inclined galaxy ($i_2=71\deg$, $\omega_2=30\deg$). The two fiducial
models employ different galaxies: in the first case (referred to as the
``disk/halo" model), the galaxies are both composite disk/halo galaxies
lacking a bulge component, while in the second case (the ``disk/bulge/halo"
model) both galaxies also possess dense bulges with a bulge:disk mass ratio
of 3:1.

\subsection{Dynamical Evolution}

Figure 1 shows the evolution of the old stellar component in the fiducial
disk/halo galaxy merger.\footnote{In what follows, we refer to the
collisionless particles originally comprising the exponential disk as the
``old stellar component" and refer to the stars formed via star formation as
the ``young stellar component."} Shortly after their initial close approach
at $t \sim 24$, the galaxies become severely distorted, forming long tidal
tails and a bridge connecting the two galaxies.  In response to the tidal
force from the passing companion, the inner regions of each disk also form
linear bar-like structures. The galaxies continue to separate after the
initial passage; however, the dark matter halos of the galaxies are
``spun up" by the encounter, due to an efficient conversion of orbital
angular momentum into internal spin (Barnes 1992; Hernquist 1992, 1993a).
As a result, the galaxies brake on their orbit, achieving a maximum
separation of $r_{apo} \sim 12$ before turning around and falling back
together.

As the galaxies encounter one another again, they experience a second quick
passage, with $r_{peri} \sim 1.5$ and $r_{apo} \sim 2.5$, before finally
merging at $t \sim 65$. Violent relaxation transforms the stellar component
into a distribution which possesses an \rquart\ law surface density profile
over a large range in radius (Barnes 1992; Hernquist 1992, 1993a).  Irregular
structure in the inner regions is rapidly mixed away, leaving a remnant which
resembles -- at least superficially -- an elliptical galaxy. At large radii
(\ie $r > 3$), where the dynamical timescale is much longer, irregular
structures such as shells, plumes, and tails persist; this ``fine structure"
may provide a long-lived signature of the merger in the remnant galaxy.

While the stellar component in each galaxy follows a collisionless
evolution, the gaseous components, shown in Figure 2, are subjected to
strong shocking and dissipation of energy and angular momentum, and
evolve in a distinctly different manner. During the initial close passage,
gas shocks at the interface between the two galaxies, and forms a physical
bridge between them as they begin to separate (see $t=28.8$). At first, the
gas in the disks responds much like the stars, but the response in the gas
is sharper as it crowds along the stellar bars. The nonaxisymmetric potential
of the distorted disks acts to torque the disk gas (see \S 3.3), driving a
rapid radial inflow of gas --- by $t=48$, approximately half of all the gas
has been driven into the inner few hundred parsecs of the galaxies, where
it fuels intense starburst activity (\S 3.2).

As the galaxies approach the final stages of merging, the rapidly changing
gravitational potential forces much of the remaining gas onto intersecting
orbits, where it shocks, dissipates energy, flows to the center
of the remnant, and forms stars. By the time the merger is complete,
approximately three quarters of the gas originally in the disks ends up
in a compact starburst core at the center of the remnant (Mihos \&
Hernquist 1994c). Most of the remaining gas has settled into a warped,
diffuse disk at intermediate radii, with a small fraction of
gas remaining at large radii in the tidal tails, where it will continue
to rain back in on the remnant over many Gyr.

To illustrate the consequences of varying the internal structure of the merging
galaxies, Figure 3 shows the evolution of the old stellar component in a
merger involving galaxies with compact central bulges. Because the
bulges comprise only $\sim$ 5\% of the total dynamical mass of the
galaxies, the global evolution of this simulation is very similar to the
fiducial disk/bulge merger. The important difference lies in the
response of the inner disk; rather than forming a strong bar after the
initial collision, the disk sports features more reminiscent of two-armed
spiral patterns (see $t=33.6$). These structures are less effective at
driving gaseous inflows in the disk than bars; thus, while gas flows
inwards in each galaxy, it does so only slowly, and settles into a
quasi-steady state before reaching the center. As a result, the distribution
of gas in the central regions of the galaxies is much more diffuse than
in the disk/halo mergers, as suggested by Figure 4; the star formation
rate is elevated only modestly during this portion of the evolution, and
the gas is not greatly depleted.

When these disk/bulge/halo galaxies finally merge, strong gravitational
torques deprive the gas of nearly all of its angular momentum, driving it
into the central regions of the remnant. In only $\sim$ 20--30 Myr, over
half the remaining gas in the system settles into a massive ``cloud"
a few hundred parsecs in diameter. The high gas densities in this cloud
fuel an intense starburst, creating a compact, young population of stars
in the center of the remnant. Although the history of dissipation and
star formation is very different in our two fiducial simulations, the net
conversion of gas into young stars is remarkably similar --- about 75\% of
the gas is depleted in each case. As for the disk/halo merger, the remaining
gas in the disk/bulge/halo remnant is distributed either in a warped gas
disk at intermediate radii, or lies at much greater distances in the tidal
tails. Again, infall of gas from the tidal features continues over long
timescales.

\subsection{Star Formation}

The star-forming response of the galaxies during the fiducial mergers
has been described elsewhere (Mihos \& Hernquist 1994b); here we
describe  this aspect of the simulations only briefly. We note again
that our star forming prescription is the density-dependent Schmidt law,
such that peaks in the modeled star formation rate indicate periods of
strong inflow and peaks in gas densities in the galaxies. While the use
of a Schmidt law in the extreme central environments of merging galaxies
may be a suspect parametrization, it should at least give an estimate of the
onset and timescale for the induced starburst activity. However, predictions
of absolute levels of star formation will carry significant uncertainties.

In the disk/halo mergers, early dissipation and inflow produces central
starbursts in the disks shortly after their initial encounter, when the
galaxies are still widely separated, as can be inferred from the star
formation rates shown in Figure 5a (see Mihos \& Hernquist 1994b for
images). The star formation rate (SFR) in each disk peak at 20--30 times
their initial amplitude, with an order-of-magnitude increase sustained for
$\sim$ 150 Myr. As these starbursts deplete the nuclear regions of gas (Figure
5b), the starbursts die out and the global SFR declines to its pre-encounter
level well before the galaxies merge.  Dissipation during the final stages
of the merger fuels a relatively weak starburst in the center of the
remnant,\footnote{We note a discrepancy between Figure 5 and the similar
Figure 2 of Mihos \& Hernquist (1994b). Because of the relatively coarse
timesteps used by Mihos \& Hernquist (1994b), the dynamics of material near
the dense cores was somewhat compromised. Our new more careful (and costly)
simulation corrects that problem; however, while some small quantitative
differences exist, the overall conclusions of Mihos \& Hernquist (1994b)
remain unchanged. We comment further on this issue in \S4.1.} which rapidly
fades.  Continued infall of gas from the tidal tails is not sufficient to
fuel significant levels of further star formation, and the remnant evolves
passively thereafter.

The merging disk/bulge/halo galaxies have a much different star-forming
history. Because the spiral response of the disks to the interaction is
much less effective at driving gaseous inflows than bars, the starbursts
which develop after the initial encounter are much weaker in strength than
those in the bulgeless galaxy mergers. The global SFR is increased by only
a factor of a few during the intermediate stages of the interaction (Figure
5a), and the disk gas is not significantly depleted (Figure 5b). Consequently,
when the galaxies finally do coalesce, they have fully twice the gas mass
remaining as their bulgeless counterparts, and the rapid collapse of this
gas drives a furious starburst at the center of the merging system. To the
extent that a Schmidt law is valid, the SFR peaks at more than $\sim$ 70
times its pre-interaction rate, and the rapid gas depletion burns the
starburst out after only $\sim$ 50 Myr. It is important to note that
the relative SFR shown in Figure 5a is defined with respect to its
pre-interaction value in the galaxy {\it pair}\ ; because the starburst
triggered in the disk/bulge/halo merger occurs in one merged object, the
factor of $\sim$ 70 increase in relative SFR corresponds to a rate which is
nearly 150 times larger than that in a {\it single} isolated disk galaxy.
Again, once this starburst dies out, the remnant evolves only passively in
terms of its star forming properties.

It is interesting to note that while the fiducial models have very
different star-forming histories, their {\it time integrated} star-forming
properties are quite similar. In both cases, approximately 75\% of the
original disk gas is converted into stars, leaving behind a dense stellar
core in the center of the remnant (Figure 6; see also Mihos \& Hernquist
1994c). While a significant fraction of gas is launched to great distances
in the tidal tails, an easily detectable amount of cold gas remains in
the remnant
($\sim$ a few $\times 10^9$ M$_{\sun}$).  This gas mostly lies in a
warped, diffuse disk, similar to the HI disk seen in the peculiar elliptical
galaxy NGC 5128 (Cen A; van Gorkom \etal 1990; Schiminovich \etal 1994).
We also note the presence of small star-forming clumps of gas and stars
in the tidal tails; while the gasdynamics of the tails is somewhat
compromised by the use of an isothermal equation of state for the gas,
star-forming objects like these have been seen in the tails of merger
candidates such as NGC 7252 (Schweizer 1982), NGC 4038/39 (Mirabel \etal
1992), Arp 105 (Duc \& Mirabel 1994), and the ``Superantennae" (IRAS
19254-7245; Mirabel \etal 1991). It seems clear from these models that
the combined effects of dissipation and star formation should leave behind
long-lived signatures of the merging process which may potentially be used
to identify present-day galaxies as products of mergers.

\subsection{Gas Inflow}

To identify the mechanisms responsible for the inflows in the merging
galaxies, we first isolate the gas which suffers the greatest loss of
angular momentum, and then decompose the time-dependent torques acting on
these particles. Owing to star formation, which converts gas into stars,
this material is more correctly a ``gas$+$young star" mass; however, since
the starburst population is born only when the inflow has mostly run
its course, these particles are predominantly gaseous during the period
of greatest loss of angular momentum. In this section, therefore, we will
speak of angular momentum and torques acting on the {\it gas}, but the
reader should bear in mind that the matter being tracked in the analysis
is actually the hybrid gas/young star particles.

The first step is to identify all gas/young star particles in the central
clump (at $r<0.1$, or \simlt 350 pc) of the merger remnant at late times.
In each fiducial model, this mass amounts to $\sim$75\% of the initial gas
content of the disks. Because we are interested primarily in the evolution
of the {\it spin} angular momentum of the gas (to identify inflows in the
disks), we exclude particles which fell in late from the tidal tails or
which were transferred to the companion galaxy at intermediate
times.\footnote{For a discussion of the evolution of the {\it total} angular
momentum of the gas, see Barnes \& Hernquist (1995).} This cut, which
removes $\sim 20$\% of the particles in the central clump, allows for a much
cleaner estimate of the spin angular momentum and torques acting on the
bulk of the gas in each galaxy. The final subset of gas particles used in
the analysis comprises 59\% (disk/bulge merger) or 56\% (disk/bulge/halo
merger) of the initial gas content of the disks.

Figure 7a shows the evolution of the spin angular momentum of
the gas in the prograde disk of the fiducial disk/halo galaxy merger.
At closest approach, $t\sim 24$, the spin increases slightly as angular
momentum is transferred from the orbital motion of the galaxies
to the internal spin of the disks, and the tidal tails are formed.
Shortly thereafter, the gas rapidly dissipates energy and loses angular
momentum, as it shocks along the bar, and flows inwards towards the
galaxy center. In the inclined disk, the response is similar, although
differences resulting from the geometry are apparent. In particular,
gas in the inclined disk loses less angular momentum than gas
in the prograde disk; however, these differences are small, and
strong inflow develops in the inclined disk as well.

These inflows rapidly transport gas to the inner regions of the disks.
By $t=48$, when the galaxies are still widely separated, $\sim$ 80\% of the
material which ultimately ends up in the remnant core has already collapsed
to within $r<0.1$ (350 pc) of its host galaxy's center. When the galaxies
merge at $t\sim 65$, these dense cores coalesce in a mostly collisionless
manner (since the gas has largely converted to young stars); a small amount
of the remaining gas is driven to the remnant center to complete the era of
strong dissipation in the merger.

To identify the factors responsible for the gas inflow in the prograde galaxy,
we decompose the torques acting on the gas.  Figure 7b shows the gravitational
and hydrodynamical torques.  The latter are always small, except during
the initial collision, when they {\it add} spin angular momentum to the gas,
as found also by Barnes \& Hernquist (1995). At later times, when the
galaxies are finally merging ($t \sim 65$), hydrodynamic torques are also
small, at least partly due to the fact that the gas has largely been
converted to stars by that time. The gravitational torques are strongest
during the period $t \sim$ 25--40, corresponding to the times of most rapid
inflow. When the galaxies merge, the rapidly varying gravitational potential
acts to strongly torque the gas/young star population, driving an additional
small inflow towards the center of the remnant.

Figure 7c further decomposes the gravitational torques into those
from the host galaxy and those from the merging companion.
The companion mainly {\it spins up} the gas on successive
passages (\ie $t\sim$ 24, 60), acting against the inflow.
In fact, the gravitational torques from the host galaxy itself
are mostly responsible for the inflow, and a Fourier analysis of the old
stellar component of the disk (Figure 7d) shows that these torques
coincide with the growth and strength of the $m=2$ mode
in the stellar disk. During the final merger, the companion acts as
both a sink and a source for angular momentum, but the host galaxy primarily
drives the inflow. At these late times, however, the modal analysis
becomes ill-defined, as violent relaxation largely destroys the disk
of the galaxy.

This analysis confirms quantitatively the picture outlined in \S3.1.
During the initial collision, gravitational torques from the companion
spin up the prograde disk, forming the tidal tails and triggering the
growth of a strong $m=2$ bar mode in the disk. The response of the stars
and gas to the perturbation is slightly different, as the gas bar
tends to lead the stellar bar by a few degrees (Barnes \& Hernquist
1991). The segregation of stars and gas along the bar allows the stellar
bar to gravitationally torque the gas and initiate a rapid inflow while
the galaxies are still widely separated. A powerful central starburst ensues,
rapidly depleting the galaxies of a large fraction of their ISM. When these
galaxies finally merge, the rapidly varying gravitational torques drive a
small additional amount of gas towards the center, fueling a weak starburst.

A similar analysis of the prograde disk in the disk/bulge/halo merger,
shown in Figure 8, provides an interesting contrast to the evolution
detailed above. Unlike the rapid, early loss of spin angular momentum
by the gas seen in the disk/halo merger, the gas in the disk/bulge/halo
merger experiences a two-stage dissipational history, as can be inferred
from Figure 8a.  The gas sheds $\sim$ 60\% of its initial spin angular
momentum shortly after first close approach, but the inflow is retarded at
$t\sim$ 35 as the inflowing gas reaches a new dynamical equilibrium. This
state is disrupted when the galaxies finally merge, triggering a second
period of strong inflow during which the massive gas cloud forms
in center of the remnant, fueling a violent starburst.

The torque decomposition shown in Figure 8b again demonstrates that
hydrodynamic torques, important only when the galaxies are physically in
contact with one another, tend to {\it add} angular momentum to the gas,
while the gravitational torques are responsible for the inflow.
The first inflow is again driven largely by the host disk; however, a
comparison of Figures 8 and 7 shows that the torques are weaker
in the disk/bulge/halo galaxies, resulting in a less violent response.
The relative weakness of these torques is directly attributable to the
milder reaction of the stellar disk to the encounter. Figure 8d shows a
Fourier analysis of the mass density of the stellar disk; while the $m=2$
mode is clearly visible, it is reduced in strength compared to that in the
fiducial disk/halo model, as the response of this disk is more in the
fashion of a two-armed spiral rather than a strong linear bar mode. When
the galaxies finally do merge, gravitational torques from both galaxies
act to drive the gas into the center of the remnant.

What causes the different responses of the stellar disks in the different
galaxy models? Figure 9 shows a plot of the disk stability parameters
$Q$ (Toomre 1962) and $X_2$ (Toomre 1981). In both models, $Q > 1$ over
the entire disk, ensuring that the disk does not suffer from local
axisymmetric instabilities. However, the $X_2$ parameter, which measures
the stability of disks to the growth of global $m=2$ modes, is markedly
different in the inner regions of the two models. In the disk/halo model,
$X_2 < 1$ over the inner scale length of the galaxy, showing that the disk
is susceptible to strong amplification of any $m=2$ perturbation.  As a
result, this disk responds more violently to the companion's tidal field,
driving the gas inwards early in the interaction. By contrast, the inclusion
of a dynamically hot bulge component in the disk/bulge/halo galaxy model
stabilizes the inner regions of the disk against the growth of $m=2$
modes, resulting in a weaker response to the perturbation of the companion.
As shown by the torque analysis, this weaker $m=2$ mode is less effective
at driving an inflow, and early starbursts are inhibited in these galaxies.

In essence, therefore, we have a description of gas inflow and starbursts
in mergers in which those effects are at least partly regulated by the
internal structure of the merging galaxies. Galaxies which lack central
bulges to stabilize them against $m=2$ structures respond violently after
the initial collision, forming a strong central bar. This bar acts to
drives gaseous inflow early in the encounter, giving rise to starburst
activity while the galaxies are still widely separated. These early
starbursts quickly deplete the gas, so that when the galaxies finally
merge, they are relatively gas poor and suffer only weak starbursts. By
contrast, the inclusion of a compact central bulge stabilizes the galaxies
against the development of strong bars and inhibits early inflow and starburst
activity. When these gas-rich galaxies merge, the strong gravitational
torques drive the gas to the center of the remnant and fuel an intense
but very short-lived starburst.

\section{Varied Geometries}

To assess the role played by encounter geometry in fueling inflow and
starburst activity we have run other models similar to the fiducial
mergers described in \S3, but with varying disk inclinations. As in
the fiducial models, these encounters all involve equal mass
galaxies colliding on initially parabolic orbits with $r_{init}=30$
and $r_{peri}=2.5$. While the fiducial models followed a prograde-inclined
collision ($i_1=0\deg , \omega_1$ undefined; $i_2=71\deg , \omega_2=30\deg$),
these additional models involve a prograde-prograde collision
($i_1=0\deg , \omega_1$ undefined; $i_2=0\deg , \omega_2$ undefined),
a prograde-retrograde collision
($i_1=0\deg , \omega_1$ undefined; $i_2=180\deg , \omega_2$ undefined),
and an inclined-inclined collision
($i_1=71\deg , \omega_1=-30\deg$ ; $i_2=71\deg , \omega_2=30\deg$).
For {\it each} choice of geometry we ran one simulation involving
two bulgeless galaxies and one simulation involving galaxies with
central bulges, with a bulge:disk mass ratio of 1:3.
While this set of eight models is not an exhaustive survey of
encounter parameters, the calculations allow us to
explore the robustness of some of our claims based on the fiducial models.

\subsection{Time Step Concerns}

Because of the great computational cost of our simulations, the extra
simulations were run without refining the timestep by a factor of
three when the dense clouds of gas developed in the galaxies, as was
done for the fiducial models of \S 3. Using the coarser timestep throughout
the simulation reduced the CPU requirements considerably.  There is a
price to pay for this compromise, however; at late times in the calculation,
when dense gaseous cores form in the galaxies' centers, the use of a coarser
timestep entails a poor integration of the orbits of particles near these
gas clumps.  In particular, energy conservation is degraded, with total
energy drifts amounting to $\sim$ 10\%.

How does this loss in computational accuracy affect the results? In large
part, the errors are confined to the inner regions of the galaxies, where
the gas clouds dominate the dynamics. For example, models which include
dense bulges experience a ``numerical scattering" of the bulge particles,
artificially lowering the central stellar density of the galaxies after
they merge. However, the {\it global} dynamics of the merger are mostly
unaffected, as we have confirmed by comparing the fiducial simulations with
and without refined timesteps. Accordingly, our claims about the global
morphology and stellar and gas kinematics seem sound.

Of greater concern are the consequences for the triggering of gas inflows
and starburst activity in the central regions of the remnant. Figure 10
shows a comparison of the star formation rates in fiducial models with and
without the finer timesteps.  For the disk/bulge/halo mergers, the differences
are always small, due to the fact that the dense cores do not form until
the final stages of the merger, and thus do not have time to strongly alter
the structure of the inner regions of the remnant. In the disk/halo
mergers, the gas clouds form earlier and can subsequently influence the
detailed evolution of the centers of the galaxies as they merge. Early
in the encounter, differences between the two disk/halo merger simulations
are small; only during the final merger is any discrepancy noticeable.
The cruder models {\it underestimate} the amount of dissipation and starburst
activity during this phase; however, these differences are minor (a factor
of at most a few in starburst intensity), and are apparent for
just a brief period, suggesting that qualitative results based on the use of
coarse timesteps are reliable.

\subsection{A Prograde-Retrograde Merger}

As an example of varying the encounter geometry, we describe in detail the
merger of two disk/bulge/halo galaxies from a prograde-retrograde orbit.
While many of the parameters of this simulation -- \ie the mass ratio,
galaxy structure, and orbital type -- are identical to the fiducial
disk/bulge/halo model shown in Figures 3 and 4, the substitution of a
retrograde disk for an inclined disk leads to some interesting differences
in the detailed evolution of the merger. First, the retrograde disk
experiences a different different response since the orbits of particles
in the disk are highly nonresonant with the orbital motion of the companion.
Second, because of the co-planar nature of this interaction, the potential
for strong shocks to develop in the colliding ISMs of the galaxies as they
interpenetrate is enhanced.

Figure 11 shows the evolution of the old disk stars during this merger.
The prograde disk evolves in largely the same manner as that shown in Figure
3, but the retrograde disk exhibits a markedly different morphology. Because
of the lack of a strong resonance between the orbital and rotational motions
of the particles in the retrograde disk, tail formation is inhibited. Instead,
the disk develops a transient, leading tidal arm in addition to a strong
two-armed spiral pattern, as can be seen at \eg $t=28.8$ in Figure 11. Even
without the extended tails, the retrograde galaxy still displays distinct
morphological features indicative of the collision: tidally-induced spiral
arms,
a central bar, and diffuse tidal debris composed of both its own stars
and some stars stripped from the prograde companion. After passing
by one another, the galaxies turn around on their orbits and merge,
somewhat more rapidly than in the fiducial model of Figure 3. The more rapid
merging is likely due to increased drag from the overlapping ISMs of the
galaxies; as the galaxies collide, the gas shocks, dissipates energy, and lags
the orbital motion of the galaxies, hastening the braking of the
galaxies on their orbits. This effect is enhanced in co-planar mergers,
as a larger fraction of the ISM in the galaxies shocks and dissipates
energy during the initial impact (see also Mihos 1992; Mihos \etal 1992).

The hybrid gas/young star component of the galaxies is shown in Figure 12,
and emphasizes even more dramatically the differences from the fiducial
encounter.  Because the retrograde disk counterrotates with respect to the
orbital motion of the companion, gas in the retrograde disk rotates into,
rather than away from, the collisional interface of the overlapping ISMs.
This situation, coupled with the fact that the collision is co-planar,
results in a stronger shock and allows the retrograde disk to sweep up
and accrete a substantial fraction ($\sim$ 30\%) of the gas in the prograde
disk. While the tidal tail of the prograde disk forms mostly out of material
traveling on non-intersecting orbits --- and is hence gas-rich --- the spiral
features in the stellar component of the retrograde disk are supported by
material which has passed through the prograde companion. Because the gas
shocks at this interface and cannot freely stream through with the tidal arms,
the arms in the retrograde disk are gas-poor. Such gas-poor tidal features
in an otherwise gas-rich interacting galaxy could therefore serve as a
strong indicator of a retrograde collision.

Finally, a map of star formation in the merging pair is shown in Figure
13. As in the fiducial encounter, star formation is boosted   in gas
initially compressed along the tidal features (see also Mihos \& Hernquist
1994b), but most stars formed at intermediate times (\ie $t=48$) come from
gas which has lost angular momentum and fallen into the inner few kpc of
the galaxies' centers. Interestingly, the star formation rate in the
retrograde disk is twice that of the prograde disk at these intermediate
times, due largely to the differing gas content of the two galaxies. As
noted above, the retrograde disk accretes $\sim$ 30\% of the gas from the
prograde disk; furthermore, the prograde disk has temporarily lost another
$\sim$ 15-20\% of its gas to its tidal tail. As a result, at this time the
retrograde galaxy contains twice the amount of gas as its prograde companion,
and this added gas fuels a higher star formation rate (see also Mihos 1994).
While the star formation intensity is still rather modest, this finding
contrasts with previous suggestions that star formation rates in interacting
galaxies will be higher in prograde than retrograde disks. When
the galaxies merge at $t\sim$ 60, the remaining gas in both disks is
rapidly driven to the center of the remnant, fueling an intense starburst
(see \S 4.3 below). After the merger is complete, some additional star-forming
gas remains in a small central disk.

\subsection{Star Forming Response}

The evolution of the global star formation rate (SFR) for each of the eight
mergers is shown in Figure 14. Again, because of the use of a density-dependent
Schmidt law to regulate star formation, Figure 14 can be seen equivalently as
tracking inflow and gas density.  For galaxies with similar structure,
the qualitative results are mostly independent of encounter geometry.
That is, in the merging disk/halo galaxies, gas inflow occurs shortly
after the initial collision, fueling starburst activity while the galaxies
are still widely separated. On the other hand, in galaxies which include
bulges, the inflow and starburst activity occur late in the merger --
regardless of geometry -- when the galaxies are coalescing. While geometry
certainly plays some role in determining the kinematic and morphological
responses of the galaxies, the structure of the galaxies appears to be the
decisive factor in terms of the induced starburst activity. Encounter
geometry may be more important in in distant encounters (\eg Mihos \etal 1992,
Mihos 1994) or in mergers of galaxies with bulges significantly less dense
than those in our fiducial model (Barnes \& Hernquist 1995).

However, while galaxy structure dominates the star forming response of
our models, encounter geometry is not entirely negligible. The most intense
inflows and starbursts develop in the co-planar prograde-prograde
or prograde-retrograde encounters, while weakest ones occur
when the galaxies are both inclined. This effect is most noticeable
in plots of the evolving gas content of the various
simulations in Figure 15, which measure the time-integrated
strength of the induced starbursts. These starbursts deplete $\sim$
65\% of the gas in the inclined disks, and as much as $\sim$ 85\% of
the gas in co-planar mergers. This trend can be explained by more
efficient dissipation in the co-planar mergers. Because gas can
easily shock and dissipate energy when the disks are co-planar, more
of the gas is driven into the central regions in these encounters,
fueling more intense starbursts and resulting in greater gas depletion. By
contrast, in inclined encounters the gas is perturbed onto orbits
which takes it out of the disk plane, reducing the chances of strong
dissipation. The gas inflows and associated starbursts
are accordingly somewhat weaker than in co-planar mergers.

A second geometrical effect can be seen in the evolution of the
merging galaxies: the onset time for the activity is slightly
different as the galaxies in the various simulation merge. In the disk/halo
models, the strongest inflows occur in the galaxies $\sim$ 15 time units
($\sim$ 200 Myr) after the initial collision, regardless of geometry, and
can be attributed to the growth of the central bar over a dynamical timescale
of the disk. However, in the disk/bulge/halo mergers the onset
of the inflow occurs as early as $t=55$ for the prograde-prograde galaxy
case and as late as $t=75$ for the merger with both galaxies inclined --
a difference of $\sim$ 250 Myr. Rather than a consequence of variations in
the dissipational history of the gas, this effect simply results from
the different coalescence timescales of the mergers. When the galaxies
are both prograde, merging occurs rapidly, while mergers of highly inclined
galaxies occur more slowly.  Therefore, while peak gas densities (and
associated starbursts) are achieved at different absolute times
(measured from the moment of first collision), they occur at roughly the
same dynamical stages of the merging process, when the galaxies are
ultimately coalescing and are within a kiloparsec or so of one another.

\section{Discussion}

\subsection{Theoretical issues}

\subsubsection{Relation to previous results}

The results of this paper reinforce and extend those of earlier
studies of gasdynamics in major mergers.  In agreement with, \eg
Negroponte \& White (1983) and Barnes \& Hernquist (1991, 1995), we
find that the tidal forces operating on disks during mergers can initiate
radial inflows of gas into the nucleus of a remnant, forcing substantial
fractions ($\sim 75 \%$) of the gas initially spread throughout both
galaxies into localized regions only a few hundred kpc across.  If a
density-dependent Schmidt law is appropriate to describe star formation
under these extreme conditions, then the densities in these central gas clumps
are sufficiently high to trigger short-lived, but powerful, bursts of star
formation with rates similar to those observed in the ultraluminous
infrared galaxies.  This inference generalizes the work of, \eg
Noguchi \& Ishibashi (1986) and Mihos \etal (1992, 1993), who
discovered that transient encounters between disk galaxies can also
induce nuclear starbursts in disks, but of considerably weaker
intensity than those described in \S 3.

\subsubsection{Inflow mechanism}

By isolating the torques acting on the gas driven to the center of a
remnant during a major merger, we have demonstrated that this gas loses
angular momentum {\it gravitationally}, primarily through its
interaction with disk stars, substantiating similar claims by, \eg
Noguchi (1987), Combes, Dupraz, \& Gerin (1990), and Barnes \&
Hernquist (1991, 1995).  We furthermore show that most of the torque
on the gas in a specific disk is contributed by the collisionless
matter in the {\it same} galaxy, rather than by the companion, through
the development of large-amplitude, non-axisymmetric features in the
stellar disk. Thus, the gas inflows analyzed here are analogous to those
provoked by minor mergers (\eg Hernquist 1989, 1991; Hernquist \&
Mihos 1995).

However, the basic mechanism responsible for the differences between
the response of the gas and the stars in a tidally-perturbed disk has
yet to be rigorously established.  As argued by Barnes \& Hernquist (1995),
it seems plausible that the dissipational nature of the gas will invariably
cause it to lose energy relative to the surrounding collisionless matter when
the gravitational potential varies rapidly in time.  But in the absence of
an underlying theory to describe this aspect of the dynamics, it is less
obvious that the gas must also lose angular momentum to the collisionless
particles.  It is somewhat reassuring, however, that a variety of independent
studies have shown that gas inflow appears to occur generically in barred or
non-axisymmetric disks (\eg Schwarz 1984, Athanassoula 1993).

\subsubsection{Influence of galaxy structure}

An interesting new finding of the current investigation is the
demonstration that the time-history of the gas inflows and the
resulting starbursts depends sensitively on the detailed structure of
the progenitor galaxies.  Collisions between disk/halo galaxies lead
to prompt inflows shortly after their first close passage, while the
gas in similar galaxies having compact bulges suffers this fate
only during the {\it final} stages of a merger.  The latter situation
appears to be more representative of observed systems which exhibit the
most intense starbursts (\eg Sanders 1992).

Superficially, our result appears to be at variance with that of
Barnes \& Hernquist (1991, 1995) whose mergers of disk/bulge/halo galaxies were
accompanied by significant gas inflows immediately after {\it first}
collision.  In fact, Barnes \& Hernquist employed galaxy models whose
bulges were significantly less concentrated than those described here.
We specifically chose parameters so that the rotation curves of our
galaxies rise sharply in their inner regions.  While we believe
that our models are a better description of spirals with relatively
massive bulges than those of Barnes \& Hernquist, a more systematic survey
of the parameter space of galaxy types is clearly needed to statistically
compare the simulations to starbursting galaxies.

\subsubsection{Influence of orbital geometry}

Perhaps somewhat surprising, the results summarized in \S 4 suggest
that the onset of activity in galaxy collisions is insensitive to the
inclinations of the merging disks. For example, the amount of gas
driven to the central regions of the galaxies (and subsequently consumed
in the starburst) varies only weakly for varying disk inclinations (see,
\eg Figure 15). Furthermore, to the extent that our model for star
formation is appropriate, the maximum star formation rates in Figure
14 differ by factors of only $\sim 2$ for the mergers involving the same
progenitors but for varying disk inclinations.  Unless a more thorough
exploration of parameter space predicts different behavior from the cases
we considered, it appears unlikely that the relatively modest differences
suggested by Figure 14 could be deduced from an observed sample of
interacting galaxies, given the more sensitive dependence of starburst
intensity to galaxy type.  The fact that Keel (1993, 1995) finds that the
triggering of starbursts and Seyfert activity in galaxy pairs is more
intimately linked to the shape of the galaxies' rotation curves, rather
than their inferred orbital geometry, lends observational support to the
theoretical results presented here.

As yet, we have not examined the effects of modifying orbit type or
impact parameter.  The simulations analyzed in \S\S 3 and 4 all
employed zero-energy orbits with small impact parameters.  Barnes \&
Hernquist (1995) also adopted zero-energy orbits in their modeling but
did compare simulations with impact parameters differing from one another
by a factor of two.  Although Barnes \& Hernquist did not include star
formation in their calculations, their results imply that the time-history of
tidally-induced starbursts will depend on impact parameter, as argued
by, \eg Mihos \etal (1992, 1993).  Whether or not this effect is as
significant as the dependence on galaxy morphology remains to be seen.

\subsection{Relation to observations}

\subsubsection{Starbursts, ultraluminous systems \& AGN}

Previous theoretical modeling has hinted at the possibility that mergers
can explain the ultraluminous infrared galaxies (\eg Barnes \& Hernquist
1991).  Our work supports these claims by elucidating the mechanisms
responsible for triggering gas inflows during galaxy collisions and,
furthermore, suggests that star formation rates compatible with
those observed are a natural consequence of the dynamics attending
these inflows.  The results described here, therefore, solidify the
apparent observational link between galaxy mergers and the onset of
activity in some peculiar galaxies.

Our modeling efforts also shed light on the discrepancy between the
merger timescales and the short gas depletion times inferred from
observations (see, \eg Larson 1987; Norman \& Scoville 1988; Scoville
\& Norman 1988; Norman 1988). In particular, previous models of
gas dynamics in merging galaxies indicated that inflows of gas
occur within a dynamical timescale ($\sim$ a few $\times 10^8$ years)
of the initial collision, well before the galaxies ultimately
merge (Barnes \& Hernquist 1991, 1995; Mihos \etal 1992, 1993).
However, observations of ultraluminous systems suggest that
the onset of activity occurs largely during the final stages of
merging (\eg Sanders \etal 1988; Murphy \etal 1995), and that the
timescale for starburst activity is very short ($10^7$ -- $10^8$ years).
Our models resolve this discrepancy by showing that the presence
of central bulges can delay the onset of activity until the
late stages of a merger. Because this result is achieved by
inhibiting the dynamical mechanisms which drive inflow, this
results is largely independent of the details of assumed Schmidt
law for star formation.

The absolute level of starburst activity achieved, however,
is more sensitive to our parameterization of the physics of star
formation. Within these uncertainties, our models indicate that star
formation rates may be elevated by at least two orders of magnitude
during mergers, as inferred in some observed systems (\eg Solomon \etal
1992). Our findings suggest that an AGN is not required to
account for the energy source in these ultraluminous systems.
If starbursts alone are responsible for these high luminosities, the implied
supernovae rate for a Miller-Scalo (1979) IMF is 1--2 yr$^{-1}$, which will
produce a bright compact radio source in the core of the merger remnant.
The radio continuum properties of ultraluminous galaxies may therefore
provide a strong observational constraint for this scenario; however,
current results are inconclusive (\eg Condon \etal 1991; Lonsdale \etal 1993).

By resolving the timescale discrepancy and achieving ultraluminous
levels of starburst activity, our disk/bulge/halo models now provide
a good description of the ultraluminous infrared galaxies. Studies
of ultraluminous systems show that the highest levels of emission
occur preferentially in very close galaxy pairs ($\Delta r \simlt $
a few kpc) or in single objects, rather than in widely separated
galaxy pairs (\eg Sanders \etal 1988a, Murphy \etal 1995). In comparison,
when the disk/bulge/halo mergers experience the strongest levels of
inflow (and associated starburst activity), they are in the final stages
of coalescence, with the galaxies deeply interpenetrating. The morphology
of the model galaxies at this point show distorted, irregular isophotes
and extended tidal tails, similar to those seen in optical (\eg Sanders
\etal 1988a, Melnick \& Mirabel 1990) and 21-cm HI (Hibbard \& Yun
1995) studies. Furthermore, the amounts of gas involved in our radial
inflows and the sizes of the central gas clumps which develop from them
are in reasonable agreement with observations (\eg Scoville \etal 1986;
Sargent \etal 1987, 1989; Sanders \etal 1991; Tinney \etal 1990;
Scoville 1992). Finally, within the uncertainties of the star formation
model, the extremely high rates of star formation, and the short ($\sim
5\times 10^7$ yr) duration of the starbursts compare well with the
star formation rates and starburst lifetimes inferred for ultraluminous
infrared galaxies (\eg Solomon \& Sage 1988).

Our simulations are less able to account for the few high luminosity,
large-separation pairs that do exist (Murphy \etal 1995).  The disk/bulge/halo
models achieve their highest levels of inflow and activity only once
they are within a few kiloparsecs of one another, not while they are
widely separated.  The disk/halo models, on the other hand, are most active
during the phase between first collision and the final merging when
they are separated by tens of kiloparsecs, but these collisions
yield weaker starbursts than their disk/bulge/halo counterparts .
However, {\it gas-rich} disk/halo galaxy mergers could conceivably
explain the few widely separated ultraluminous pairs. In fact,
it is likely that the progenitor galaxies which develop ultraluminous
activity possess a range of structural properties and gas contents, and
interact under a variety of conditions. Given the strong relationship between
structural properties and the timescale and intensity of starbursts in our
models, it seems plausible that a properly selected mixture of progenitor
galaxies with a range of morphological types could yield a close match to
the entire ultraluminous galaxy sample.

It is also interesting that the remnants produced in our simulations
exhibit morphological peculiarities similar to many of the brightest
radio galaxies (\eg Heckman \etal 1985, 1986; Baum \etal 1988;
Baum \& Heckman 1989a,b; Smith \etal 1990) and even to some low-redshift
quasars (\eg Stockton 1978; Heckman \etal 1984; MacKenty \& Stockton 1984;
Smith \etal 1986; Hutchings 1987; Vader \etal 1987).  At present, however,
we are unable to clarify the possible connection between galaxy interactions
and the origin of AGN, owing to the limited dynamic range of the simulations.
Nevertheless, it is plausible that continued fragmentation and collapse of
gas in the nuclei of objects like those modeled here will evolve into sources
resembling AGN (\eg Begelman \etal 1984; Shlosman \etal 1989).

\subsubsection{Ordinary galaxies}

With two possible exceptions, discussed below, our calculations
support the notion that major mergers of disk galaxies produce objects
resembling ellipticals.  In particular, loosely-bound debris falling
into the mostly-relaxed remnant produces fine-structure like that seen commonly
in ordinary ellipticals, such as shells, which were once thought to
originate mainly via minor accretion events (\eg Quinn 1984;
Hernquist \& Quinn 1988, 1989).  Similarly, as emphasized by Barnes \&
Hernquist (1995), warped disks like those discovered in galaxies such
as NGC 4753 (Steiman-Cameron \etal 1992) and NGC 5128
(Nicholson \etal 1992) are a natural consequence of the late
infall of gas into the remnant from material in the tidal tails.
Aside from these warped HI disks, such infall can
also manifest itself in the form of diffuse ``loops" of HI at
larger radii, similar to those recently observed in the peculiar
ellipticals NGC 5128 and NGC 2865 (Schminovich \etal 1994, 1995).

Star formation has the virtue of depleting the gas so that the amount
of cold gas left in the remnants are in better agreement with normal
ellipticals than in the models of, \eg Barnes \& Hernquist (1995).
The gas driven to the centers of the remnants is efficiently consumed
by the starbursts, leaving behind dense stellar cores and using up most
of the gas in each progenitor.  However, much of
the gas which falls into the remnant at late stages in a merger does not
accrete into the center but populates an extended disk, as noted
above.  In the cases we examined, typically a few $\times\ 10^9
M_{\sun}$ worth of cold gas remains, which exhibits no evidence for
vigorous star formation.  Limits on the total amount of cold gas in
ordinary ellipticals are not severe but our results appear to be in
mild conflict with present observational knowledge of normal ellipticals
(\eg Bregman, Hogg, \& Roberts 1992; see also the recent review by Roberts
\& Haynes 1994).

Perhaps more worrisome are the stellar residues of the nuclear
starbursts.  As Mihos \& Hernquist (1994c) show, the light profile of
the starburst population does not join ``seamlessly'' onto that of the
old stars in the remnant, but is instead manifest as a luminosity
``spike,'' in apparent disagreement with the core properties of typical
massive ellipticals (\eg Lauer \etal 1995).  What is the
significance of this result for the merger hypothesis?  If it is borne
out by further modeling over a wider portion of parameter space and by
exploring other descriptions of star formation, it might well rule
out the possibility that most ellipticals originate by major mergers
of pairs of disk galaxies.  Any such claim based on the limited set of
calculations done to date is, however, clearly premature, as it is not
difficult to imagine mechanisms that could ``smooth out'' these
central spikes.  As argued by Barnes \& Hernquist (1995), if the
stellar cores were subject to a good deal of violent relaxation before
finally coalescing into the body of the remnant, their luminosity
profiles would presumably be less distinguishable from the old stellar
population than in our models.  Such an effect might occur for
certain parameterizations of star formation, or if remnants like the ones
analyzed here were subject to additional merging (for discussions, see, \eg
Hernquist 1993c; Weil \& Hernquist 1995).

\subsection{Loose ends}

\subsubsection{Microphysics}

Given the approximate nature of our description of the ISM,
calculations like those discussed here are, at best,
caricatures of how the gas in colliding galaxies would actually behave.
As emphasized by Barnes \& Hernquist (1993), the use of a simple
equation of state, whether an isothermal one, as in our simulations,
or an ideal gas law combined with non-adiabatic heating and cooling,
as employed by Barnes \& Hernquist (1995), is not entirely appropriate
given that cosmic ray and magnetic pressure are probably not
negligible.  A number of authors have argued that these effects, in
particular, may enhance the pressures in the centers of galaxies by
large factors over that in the local ISM (\eg Spergel \& Blitz
1992; Heckman \etal 1990; Suchkov \etal 1993).
Nevertheless, we are encouraged by the similarity of the various
calculations to date which have used different equations of state and
treatments of radiative effects.

Perhaps more questionable is our neglect of the multiphase structure
of the ISM; a limitation necessitated by the relatively poor
resolution of the simulations.  An SPH description approximates the
ISM as a smooth fluid and, consequently, the gas cannot interpenetrate
on small scales as would an ensemble of clouds whose
collisional mean-free paths are relatively long.  Averaged over
sufficiently large scales, however, it is not implausible that such an
ensemble would be well-represented by a continuum fluid, and we are
again encouraged by the rather good agreement between our results and
those which have been obtained using a discrete-cloud model for the
ISM (\eg Negroponte \& White 1983).

\subsubsection{Star formation processes}

Given the lack of resolution on small scales, it is obvious that any
treatment of the effects of star formation will be crude and should
rightly be viewed with some skepticism.  Our parameterization of this
process is simple and yields reasonable behavior for the gas in
isolated disks over long timescales (Mihos \& Hernquist 1994e).
However, it is not at all clear that the physical conditions in
star-forming regions of quiescent disks are similar to those in the
nucleus of a starbursting galaxy.  The appropriateness of a
Miller-Scalo IMF is somewhat dubious in light of observations
suggesting that nuclear starbursts are unusually efficient at
producing high-mass stars (\eg Rieke \etal 1980; Doane \& Mathews 1993;
Doyon, Joseph, \& Wright 1994). Changes in the IMF in our
models could significantly alter the amount of gas and
energy returned from evolving stars, and subsequently impact
further star formation. A top-heavy IMF could also modify our prediction
for the luminosity profile of the merger core, produce large
amounts of hot gas from supernovae, and strongly influence the evolving
spectrophotometric properties of the remnant by changing both the metallicity
and mass function of the starburst population. Our models should clearly be
viewed as being exploratory, insofar as the IMF in starburst galaxies is still
poorly constrained.

The best scheme for incorporating feedback into the calculations is
also problematic.  As discussed in \S 2.2, energy is injected to the
surrounding gas by newly formed stars in our models through radial impulses
delivered to nearby SPH particles.  Other workers have experimented with
pure thermal input to the gas (\eg Katz 1992, Summers 1993).  Based on
empirical tests (Mihos \& Hernquist 1994e), we believe that our treatment of
feedback is well-motivated for our application, but simulations of
gravitational collapse exhibit sensitivity to the details of the feedback
mechanism (see, \eg Katz \& Gunn 1991, Navarro \& White 1993). In particular,
the rates of inflow and detailed structure of the central starburst
population may be modified by employing other schemes for energy release.
As such, the lack of a physically self-consistent treatment for handling
this feedback is perhaps the weakest aspect of our modeling technique.

\subsection{Future Directions}

Although our calculations have clarified the relevance of gas physics
and star formation to the onset of peculiar activity in the nuclei of
some galaxies, a number of fundamental questions remain unanswered.
Some starburst galaxies are accompanied by outflows (``superwinds'')
that appear to be linked to the processes responsible for the infrared
emission from these systems (\eg Baan \etal 1989; Heckman \etal 1990).
Thus far, no such winds have developed in any of the relevant numerical
simulations and their nature remains a mystery.

The possibility that mergers may trigger the formation of large
numbers of star clusters, as may be required to reconcile the merger
hypothesis with the specific frequency of globulars around ellipticals
versus spirals (\eg Schweizer 1982; Ashman \& Zepf 1992;
Whitmore \etal 1993), remains unaddressed.  Certainly, our models
show elevated star formation rates in the collisionally
shocked gas, creating an environment which could be very conducive to forming
globular clusters (\eg Murray \& Lin 1989, 1992). However, simulations with
much larger dynamic range than those presented here will be required
to examine this issue in any detail.

We have also called into question the role of major mergers of
gas-rich disks to the origin of normal elliptical galaxies.  While
possibly reconciling the central phase-space densities of spiral
progenitors with those of remnant ellipticals, dissipation in the gas
may in fact be overly efficient and yield stellar cores that are unlike
those of most large ellipticals.  Whether or not this result is of
basic importance to the origin of early-type galaxies or is instead
an artifact of the approximations employed in the simulations is
uncertain. If this finding is verified by more refined calculations,
it may provide a strong constraint on the fraction of normal ellipticals
which formed through mergers of gas-rich disk galaxies.

Finally, the simulations described here explore many different
aspects of galaxy evolution, as they follow disk galaxy collisions
through the merging process, forming intense starburst activity,
and leaving behind a remnant which appears strikingly similar to many
elliptical galaxies. For the first time, the models tie together in a
well-defined manner galaxy mergers, ultraluminous infrared galaxies, and the
likely formation history of at least some fraction of the
elliptical galaxy population of the local Universe. Because of
uncertainties in incorporating star formation physics into the models,
simulations of the processes discussed here are rightly in their infancy.
In the future, refined modeling techniques and higher resolution calculations
will make it possible to explore galaxy mergers and starbursts
in ever-greater detail, and place strong constraints on the role
played by mergers in driving the evolution of galaxy populations.

\acknowledgements

We thank Josh Barnes, Fran\c cois Schweizer, and Alar Toomre for helpful
discussions.  This work was supported in part by the San Diego Supercomputing
Center, the Pittsburgh Supercomputing Center, the Alfred P. Sloan
Foundation, NASA Theory Grant NAGW--2422, the NSF under Grants
AST 90--18526 and ASC 93--18185 and the Presidential Faculty
Fellows Program. J.C.M. is supported by NASA through a Hubble Fellowship
grant \#~HF-01074.01-94A awarded by the Space Telescope Science Institute,
which is operated by the Association of University for Research in Astronomy,
Inc., for NASA under contract NAS 5-26555.

\clearpage

\begin{figure}
\caption{Evolution of the old stellar disk component in the fiducial
disk/halo merger. The prograde disk enters from the upper right of the
frames, while the inclined disk approaches from the lower left. Each
frame measures 20 units on an edge; time is shown
in the upper right corner of each frame.}
\end{figure}

\begin{figure}
\caption{Evolution of the gas and young stellar components in the fiducial
disk/halo merger. The scales are identical to those in Figure 1.}
\end{figure}

\begin{figure}
\caption{Evolution of the old stellar disk component in the fiducial
disk/bulge/halo merger. Scales as in Figure 1.}
\end{figure}

\begin{figure}
\caption{Evolution of the gas and young stellar components in the fiducial
disk/bulge/halo merger. Scales as in Figure 1.}
\end{figure}

\begin{figure}
\caption{a) Evolution of the global star formation rate (relative to two
isolated disks) for the fiducial models. b) Evolution of the total gas
mass for the fiducial models.}
\end{figure}

\begin{figure}
\caption{A comparison of the initial and final cumulative mass distributions
of gas and young stars in the fiducial mergers. Note the compact starburst
cores and the increasing amount of gas at larger radii in the merger remnants.}
\end{figure}

\begin{figure}
\caption{Angular momentum and torque decomposition for the gas in the
prograde disk/halo galaxy. a) Spin angular momentum. b) Gravitational
and hydrodynamical torques. c) Gravitational torques from each
galaxy. d) Fourier modal analysis of the stellar prograde disk.}
\end{figure}

\begin{figure}
\caption{Angular momentum and torque decomposition for the gas in the
prograde disk/bulge/halo galaxy. a) Spin angular momentum. b) Gravitational
and hydrodynamical torques. c) Gravitational torques from each
galaxy. d) Fourier modal analysis of the stellar prograde disk.}
\end{figure}

\begin{figure}
\caption{Disk stability parameters Q and X$_2$ in the disk/halo
galaxy model (solid curves) and disk/bulge/halo galaxy model (dashed
curves).}
\end{figure}

\begin{figure}
\caption{A comparison of the star formation rates in the fiducial models
with and without the use of refined time steps. Solid lines show the
response of the models which employ refined time steps, while dotted
lines show the response of the models with the coarser time steps.}
\end{figure}

\begin{figure}
\caption{Evolution of the old stellar disk component in the
prograde-retrograde disk/halo model. The prograde disk enters from the
upper right of the frames, while the retrograde disk approaches from the
lower left. Scales as in Figure 1.}
\end{figure}

\begin{figure}
\caption{Evolution of the gas and young stellar components in the
prograde-retrograde disk/halo model. Scales as in Figure 1.}
\end{figure}

\begin{figure}
\caption{Evolution of the star forming morphology of the
prograde-retrograde disk/halo model. Because of the use of a density-dependent
Schmidt law, this Figure is comparable to a map of local gas density.
A logarithmic intensity map is used to best show faint detail. Scales as
in Figure 1.}
\end{figure}

\begin{figure}
\caption{Evolution of the global star formation rate for the various
model runs. Left panels: Disk/halo galaxy mergers. Right panels:
Disk/bulge/halo galaxy mergers. Star formation rates are measured
relative to the total star formation rate in {\it two} isolated disks.}
\end{figure}

\begin{figure}
\caption{Evolution of the total gas mass in the various simulations.
Top: Disk/halo galaxy mergers. Bottom: Disk/bulge/halo galaxy mergers.}
\end{figure}

\end{document}